# GEOMETRIC DEEP LEARNING FOR AUTOMATED LANDMARKING OF MAXILLARY ARCHES ON 3D ORAL SCANS FROM NEWBORNS WITH CLEFT LIP AND PALATE

Artur Agaronyan[a,b], HyeRan Choo[a,d,‡], Marius George Linguraru[b,c], Syed Muhammad Anwar[b,c‡],

[a]Department of Surgery, Division of Plastic and Reconstructive Surgery, Neonatal and Pediatric Craniofacial Airway Orthodontics, Stanford University School of Medicine, 770 Welch Road, Palo Alto, California, 94394, U.S.A.
[b]Sheikh Zayed Institute, Children's National Hospital, 111 Michigan Ave NW, Washington, DC 20010, U.S.A.
[c]School of Medicine and Health Sciences, George Washington University, Washington, DC 20052, U.S.A
[d]Lucile Packard Children's Hospital Stanford; 730 Welch Road, Palo Alto, California, 94304, U.S.A.

‡ Co-corresponding authors

**ABSTRACT**

Rapid advances in 3D model scanning have enabled the mass digitization of dental clay models. However, most clinicians and researchers continue to use manual morphometric analysis methods on these models such as landmarking. This is a significant step in treatment planning forcraniomaxillofacial conditions. We aimed to develop and test a geometric deep learning model that would accurately and reliably label landmarks on a complicated and specialized patient population- infants, as accurately as a human specialist without a large amount of training data. Our developed pipeline demonstrated an accuracy of 94.44% with an absolute mean error of 1.676 ± 0.959 mm on a set of 100 models acquired from newborn babies with cleft lip and palate. Our proposed pipeline has the potential to serve as a fast, accurate, and reliable quantifier of maxillary arch morphometric features, as well as an integral step towards a future fully automated dental treatment pipeline.

## 1. INTRODUCTION

Morphometric analysis of anatomical landmarks allows clinicians and researchers to identify clinically significant patterns and trends between healthy and control groups. Rapid progress in the field of intraoral 3D scanning technology has enabled the ability to generate 3D models of structures such as maxillary arches in patients with a level of precision previously unattainable [1]. These 3D meshes are much easier and faster to annotate than traditional physical cast methods. Morphometric analysis of these landmarked models can be used to predict features such as future growth patterns by identifying height of contour, discern dental health issues such as malocclusion, occlusal discrepancies, and prosthodontic restoration issues. These 3D meshes are much easier and faster to annotate than traditional physical cast methods. Morphometric analysis of these landmarked models can be used to predict features such as future growth patterns by identifying height of contour, discern dental health issues such as malocclusion, occlusal discrepancies, and prosthodontic restoration issues [2][3]. Such analysis can further aid in tailoring specialized and personalized treatment plans including alveolar moulding plates (AMP) for each patient. Such analysis can further aid in tailoring specialized and personalized treatment plans including alveolar molding plates (AMP) for each patient [4][5].

However, analysis of these scans in the clinical setting is still overwhelmingly done manually or semi-automatically. This process is time-consuming and can be prone to inconsistencies, limiting the precision of patient assessments. Smaller and specialized populations, such as infants with cleft lip and palate (CLP), also have fewer scans collected of their maxillary arches, resulting in limited data availability—a potential limitation when training current deep learning models. A properly designed deep learning (DL)-based pipeline and automation could help bridge this gap by providing consistent, precise, and rapid landmark detection, allowing clinicians to better assess growth patterns and surgical outcomes in CLP patients. For infants with CLP, accurate and timely analysis of maxillary arches is crucial for planning personalized interventions, which could ultimately lead to improved patient outcomes. Automation in analysis would not only enhance clinical decision-making but also expand data accessibility by enabling DL models to generalize effectively with smaller datasets, thereby supporting advancements in CLP research and treatment strategies.

Despite the enhanced visualizations that digital 3D scans provide compared to their traditional counterparts, the identification and annotation of anatomical landmarks on 3D models in clinical practice continue to remain largely manual tasks. This manual approach demands significant time, specialized expertise, and introduces variability and human bias, limiting the efficacy and consistency with which clinicians can leverage this 3D data. Automated detection of morphological landmarks is especially valuable in pediatric cases involving CLP, where early and frequent evaluations are necessary to assess growth, determine the need for surgical interventions, and monitor the success of treatment protocols. Automated landmark detection not only standardizes measurements across cases but also has the potential to facilitate large-scale studies, as it eliminates observer bias and increases the reproducibility of measurements. The shift from manual to automated methods could also support clinical workflows by providing practitioners with rapid, consistent, and objective information on the spatial relationships within the craniofacial region.

Deep learning approaches for 3D meshes have created a new subfield of machine learning called geometric deep learning (GDL). Traditional machine learning methods that focus on data such as images and audio excel in learning hierarchical features from the data, allowing the model to easily recognize edges, shapes, and complex patterns forming recognizable features such as a human face or an animal. However, this type of model cannot take non-Euclidean data as input, such as graphs or 3D meshes. GDL, on the other hand, adapts convolutional neural networks (CNNs) to operate on non-Euclidean structures by defining convolutions and pooling operations on graph-based representations of data, like point clouds or 3D surfaces. By doing so, GDL can capture spatial relationships and intrinsic properties of 3D shapes, allowing the model to learn and generalize geometric features directly from the mesh data.

We leverage these advances to develop a pipeline for automating the landmarking process. We particularly focus on the infant population most often neglected in the literature. In this paper, we utilize the feature point methodology proposed by Paulsen [6][7]. A 3D mesh is split into multiple views, and a feature point from one direction is seen as a ray. Using a consensus from rays of multiple 2D views, a 3D landmark can be estimated. Past attempts to landmark dental arch models with deep learning, such as Croquet [8], identified the key advantage of high repeatability and good accuracy, but do not employ tailored GDL models and were usually limited to a small number of landmarks. The contributions of our proposed pipeline are:

- A large reduction in the time required to manually landmark dental features using the automated deep learning pipeline
- We achieve higher accuracy and reproducibility than with human operators
- Our strategy requires relatively low amount of training data (<100 samples) and achieve high quality and clinically acceptable results

## 2. METHODS

Our proposed pipeline seeks to bridge the gap between technological advancements in 3D scanning and their practical application in healthcare, laying the groundwork for fully integrated, digital solutions in the evaluation and management of CLP and other craniofacial conditions. By optimizing the pipeline to handle small, delicate anatomical landmarks within infant maxillary arches, this study contributes a scalable and adaptable solution to clinical workflows. The integration of automated landmarking into the clinical setting offers not only improved consistency but also the potential for tracking of craniofacial changes, which is particularly valuable in pediatric cases where growth patterns vary widely. We aim to evaluate the effectiveness of our specialized convolutional neural network (CNN) model for automated landmarking of clinically significant landmarks (16 in this study) on a highly specialized and varied patient population with high accuracy and repeatability.

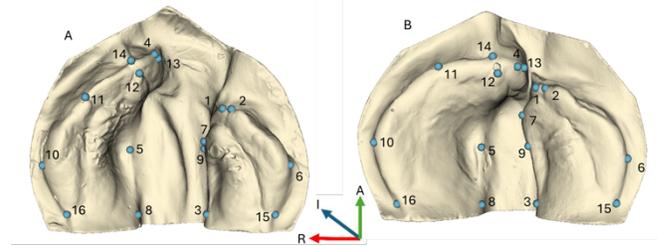

**Fig. 1.** Labeled dental landmarks on the maxillary arch, with (A) representing pre-treatment and (B) representing post-treatment. The landmarks include; Small: (1) Anterior Point, (2) Canine Point, (3) Posterior Cleft Point, (6) Widest Point, (9) Anterior Cleft Point; Large: (4) Anterior Point, (5) Anterior Cleft Point, (8) Posterior Cleft Point, (10) Widest Point, (11) Canine Point; (12) Incisive Papilla, (13) Convex Point, (14) Frenum Point, (15) Tuberosity Point, (16) Tuberosity Point.

### 2.1 Data

A total of 50 infants with unilateral cleft lip and palate (UCLP) had their maxillary arches modeled with clay models at Lucile Packard Children's Hospital Stanford/Stanford University School of Medicine under an IRB approved study. These patients then underwent BioAMP, a nasoalveolar molding therapy that gradually reshapes the gums, lip, and nose before surgery to improve surgical outcomes [9]. After treatment, these patients had their maxillary arches modeled again. These models were then scanned in a 3D model scanner and digitized into standard stereolithography (STL) models. In case of misalignment between the models, a team of dental clinicians and experts then reoriented each model to standard radiographic orientation. Each model had three planes generated, one sagittal plane intersecting the midpoint of the tuberosity line, one coronal plane intersecting the larger segment tuberosity point, and one transverse plane intersecting the larger segment tuberosity point for ease of annotation. Then, 16 clinically significant landmarks were labeled by an experienced clinician using 3D Slicer [10], including the tuberosity points, canine points, greater and lesser segments, and incisive points, on both pre- and post-treatment models for each patient [11]. Figure 1 shows these landmarks on both pre- and post-treatment models.

### 2.2 Model Training

The landmarked 3D models served as ground truth for model training. Each model was rendered into multiple 2D views, and the corresponding 3D training landmarks were projected onto these 2D images. For each 2D view, a two-stack hourglass network processes the input, which initially goes through feature extraction with residual blocks, each containing three 3x3 convolutional layers followed by batch normalization. Dropout layers are used to mitigate overfitting. The network produces 16 heatmaps, with each heatmap corresponding to one of the 16 landmarks. Each 2D image undergoes a series of stages: initial feature extraction, down sampling, intermediate refinement, and final up sampling to predict accurate landmark positions. Figure 2 illustrates the overall flow of the pipeline. The model was trained on 90 models, equally balanced between pre- and post-UCLP treatment. After training, the model was applied to a testing set consisting of 10 previously unseen intraoral models, also equally balanced. The accuracy of the model's predictions was assessed by comparing the predicted landmarks to the manually labeled ground truth in the test set. Metrics such as mean absolute error (MAE), normalized mean error (NME), and accuracy were used to quantify the performance. Accuracy was calculated by:

$$Accuracy(\%) = \left(\frac{\sum_{i=1}^{N}(e_i \leq T)}{N}\right) * 100 \quad (1)$$

where N is the total number of landmarks, $e_i$ is the error for the i-th landmark, and T is the threshold of error acceptable chosen relative to the model volume (we chose 0.01%). Additionally, manual review was conducted by a trained specialist to ensure that the predicted landmarks aligned well with clinical expectations. Training the model was performed on an NVIDIA RTX A5000 GPU, averaging 8 hours on the total dataset of 90 models. Applying the model to the test set of 10 models took 6 minutes.

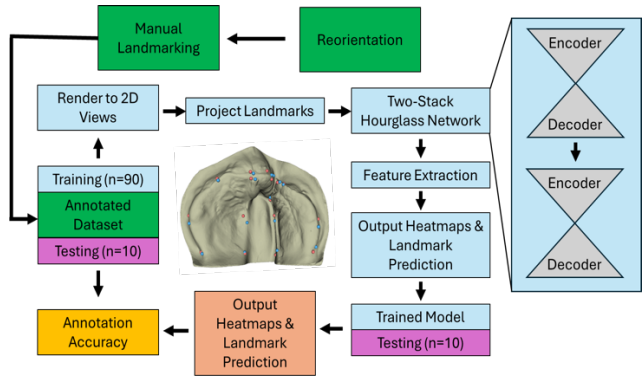

**Fig. 2.** Overview of the pipeline structure. Blue landmarks were placed manually, red landmarks were generated by the model.

## 3. RESULTS

Figure 3 shows a landmarked maxillary arch model with absolute error for each landmark. Table 1 lists error statistics per landmark for the test set (n=10). The model's performance achieved an accuracy of 94.44% considering a threshold of 0.01% of the model's volume, which is 2.56 mm, with an absolute mean error of 1.676 ± 0.959 mm. Maximum and minimum error were 6.360 mm and 0.067 mm, respectively. Skewness was 1.32. Mean volume and surface area of the sample set were 25551.697 ± 3971.654 mm$^3$ and 4480.336 ± 483.043 mm$^2$, respectively. Mean error-to-volume ratio was <0.001.

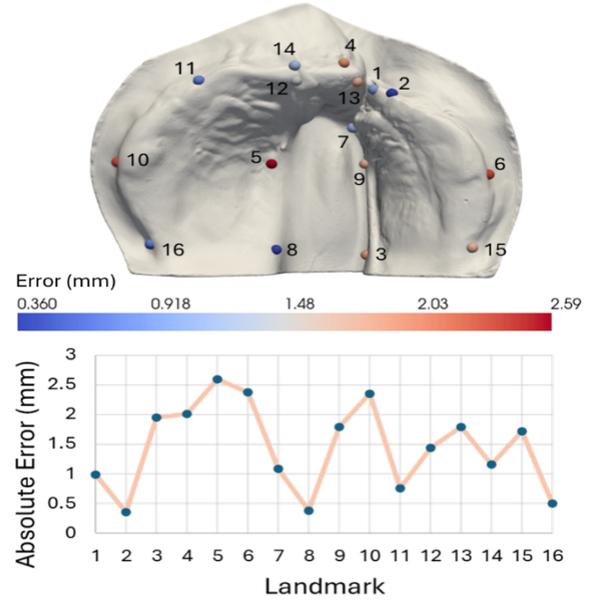

**Fig. 3.** A landmarked infant maxillary arch with 16 points. The color coding is the absolute error compared to the manually annotated landmarks.

Our pipeline's high accuracy of 94.44% indicates strong applicability to our clinical use case. The error distribution skewness of 1.32 suggests that most errors were lower, with a few higher outliers. Errors were frequently observed at the edges of the greater and lesser segments, particularly around landmarks 5, 6, 9, and 10 (see Fig. 1). This is likely due to the smoothness of these regions, which lack distinct features—such as the anterior cleft point (landmark 5), widest point (landmark 6), anterior cleft point (landmark 9), and widest point (landmark 10)—making accurate model identification challenging. Additionally, the model performed less accurately on outlier arch volumes, particularly for infants with very large or small arches.

The one-sample t-test results against a mean error of 0 for each landmark revealed statistically significant t-statistics for several landmarks, with landmark 6 (widest point) showing a t-statistic of 8.36 (p = 0.000032) and landmark 14 (frenum point) showing a t-statistic of 9.22 (p = 0.000016), indicating that these landmarks' errors differ significantly from zero. In contrast, the two-sample t-test comparing segment edge landmarks with the other landmarks yielded a t-statistic of 1.81 (p = 0.074), suggesting no significant difference in error between these groups, although the p-value is close to the threshold for significance. After applying the Bonferroni correction for multiple comparisons, the statistical significance remains unchanged.

Table 1. Error comparison of the automatically generated landmarks.

| Land mark | Root Mean Squared Error ± SD (mm) | Maximum – Minimum (mm) | Normalized Mean Error | Coefficient of Variation | Skewness |
|---|---|---|---|---|---|
| 1 | 1.42 ± 0.60 | 2.78 – 0.66 | 4.82 | 0.46 | 1.73 |
| 2 | 2.84 ± 2.03 | 6.36 – 0.36 | 7.74 | 0.97 | 1.02 |
| 3 | 1.16 ± 0.51 | 1.94 – 0.07 | 3.91 | 0.48 | -0.23 |
| 4 | 1.59 ± 0.79 | 2.38 – 0.50 | 5.18 | 0.56 | 0.16 |
| 5 | 1.69 ± 0.72 | 2.70 – 0.44 | 5.71 | 0.46 | -0.01 |
| 6 | 1.51 ± 0.51 | 2.30 – 0.78 | 5.28 | 0.36 | 0.46 |
| 7 | 2.48 ± 1.12 | 3.78 – 0.63 | 8.26 | 0.50 | -0.29 |
| 8 | 1.92 ± 1.00 | 3.65 – 0.31 | 6.17 | 0.59 | 0.62 |
| 9 | 1.97 ± 0.67 | 2.91 – 1.09 | 6.86 | 0.36 | 0.32 |
| 10 | 2.06 ± 1.09 | 3.58 – 0.38 | 6.58 | 0.61 | 0.58 |
| 11 | 2.24 ± 0.72 | 3.56 – 0.89 | 7.86 | 0.34 | 0.36 |
| 12 | 1.94 ± 0.93 | 3.22 – 0.54 | 6.39 | 0.53 | 0.27 |
| 13 | 2.20 ± 1.10 | 3.71 – 0.47 | 7.16 | 0.56 | 0.34 |
| 14 | 1.56 ± 0.49 | 2.17 – 0.49 | 5.51 | 0.33 | -0.01 |
| 15 | 2.34 ± 1.51 | 5.31 – 1.51 | 6.85 | 0.81 | 1.31 |
| 16 | 1.68 ± 0.67 | 2.20 – 1.56 | 5.73 | 0.43 | -0.87 |

## 4. DISCUSSION

Manual landmarking, while highly accurate, is time and labor-intensive. To finish landmarking one model, it took an average of approximately 45 minutes. Deep learning models reliably learn from the training dataset, identifying the relevant features most crucial to accurate labeling, and store them reliably with no loss of information, avoiding the errors human operators may be prone to, especially after repeatedly working on datasets for long hours.

Our pipeline has multiple applications in clinical use, such as facilitating large-scale studies of anatomical variations in long-term treatment studies, such as with surgery for CLP or Pierre Robin sequence. Similarly, morphological shape descriptors including palatal width and depth may be extracted with ease from a massive dataset, reducing the workload from a team of specialists working for a year to a matter of minutes and a technician that does not have to be a specialist [12]. The anatomical landmarks generated may be used for complementary measures such as surface area, volume, design of models based on maxillary arch features, curvature, and many others.

Geometric deep learning is a relatively new field of study and has received little attention for real-life applications such as clinical dentistry. However, this field holds great promise for clinical dentistry, because 3D meshes can represent many relevant features, such as tooth shape, facilitating modeling of insertion processes, and clinically significant distances between dental anatomical landmarks. For example, height of contour can inform the degree to which a tooth is tilted. This will determine what kind of treatment the tooth will need. Our model can be trained and applied for a different clinical problem in the manner described here to quickly and accurately identify the landmarks relevant to this measure and predict the type of treatment required for all the patients a clinic has, provided 3D meshes are available, without the costly and time-consuming manual analysis done by human specialists.

Similarly, landmarking is being used in pioneer studies to classify treatment outcome based on geometric relationships [13]. For example, with CLP certain geometric relationships inform the probability of a good treatment outcome before any surgery has taken place. In these studies, landmarking has been manual or semi-automated. Automated landmarking would vastly simplify this process and enable the integration of a much larger ground truth pool of patients to increase the quality of prediction and subsequent care.

## 5. CONCLUSION

This study presents a pioneering automated pipeline for accurately predicting morphological landmarks on maxillary arches, specifically designed to be accessible for non-specialists. Our results demonstrate high accuracy across a diverse test set of infants with cleft lip and palate, highlighting the effectiveness of integrating geometric deep learning in clinical applications. The implications of this work extend beyond pediatric populations, with promising avenues for future research that include adapting the pipeline for elderly and adolescent populations, functioning as a base of comparison for later non-Euclidian models that operate purely in 3D space, and integrating this technology into fully digital dental workflows. This advancement has the potential to transform dental practices by streamlining processes, reducing the need for specialist intervention, and ultimately improving patient care through enhanced predictive capabilities.

## 6. ACKNOWLEDGMENTS

This work was supported by Stanford Maternal & Child Health Research Institute, Transdisciplinary Initiatives Program [Grant number 292590 (HC)].